# Hamiltonian Formulation of Classical Fields with Fractional Derivatives

**A. A. Diab[1], R. S. Hijjawi[2], J. H. Asad[3], and J. M. Khalifeh[4]**
[1]General Studies Department- Yanbu Industrial College- P. O Box 30436- Yanbu Industrial City- KSA
[2]Dep. of Physics- Mutah University- Karak- Jordan
[3]Dep. of Physics- Tabuk University- P. O Box 741- Tabuk 71491- Saudi Arabia
[4]Dep. of Physics-Jordan University- Amman 11942 - Jordan

## Abstract

An investigation of classical fields with fractional derivatives is presented using the fractional Hamiltonian formulation. The fractional Hamilton's equations are obtained for two classical field examples. The formulation presented and the resulting equations are very similar to those appearing in classical field theory.

## 1. Introduction

Fractional calculus is an extension of classical calculus. In this branch of mathematics, definitions are established for integrals and derivatives of arbitrary non-integer (even complex) order, such as $\frac{d^{1/2} f(t)}{dt^{1/2}}$. It has started in 1695 when Leibniz presented his analysis of the derivative of order $1/2$. Then it is developed mainly as a theoretical aspect of mathematics, being considered by some of the great names in mathematics such as Euler, Lagrange, and Fourier, among many others. This branch of mathematics has experienced a revival of interest and has been used in very diverse topics such as fractal theory [1]viscoelasticty [2], electrodynamics [3,4], optics [5,6], and thermodynamics [7]. The literature of fractional calculus started with Leibniz and today is growing rapidly[8-14].

Recently fractional derivatives have played a significant role in physics, mathematics, engineering, and pure and applied mathematics in recent years[14-20]. Several attempts have been made to include non conservative forces in the Lagrangian and the Hamiltonian mechanics. Riewe[20,21] presented a new approach to mechanics that allows one to obtain the equations for non conservative systems using fractional derivatives. Eqab et al[22-24] developed a general formula for determining the potentials of arbitrary forces, conservative and non conservative, using the Laplace transform of fractional integrals and fractional derivatives.

Recently, they[24] also constructed the Hamiltonian formulation of discrete and continuous fields in terms of fractional derivatives. The fractional Hamiltonian is not uniquely defined; it seems that there are several choices of fractional Hamiltonian giving the same classical limit, i.e., the same classical Hamiltonian. This present paper is a generalization of the above work on Hamilton's equations for classical fields with Riemann-Liouville fractional derivatives.

This paper is organized as follows: In Sec.2, the definitions of fractional derivatives are discussed briefly. In Sec.3, the fractional form of Euler-Lagrangian equation in terms of functional derivative of the full Lagrangian is presented. In Sec.4, the fractional form of Euler-Lagrange equation in terms of momentum density is investigated. Section 5 is devoted to the equations of motion in terms of Hamiltonian density in fractional form. In section 6, we introduce two examples of classical fields leading to Schrödinger equation and Dirac equation in fractional derivative forms. The work closes with some concluding remarks (section 7).

## 2. Definitions of Fractional Derivatives

In this section two different definitions of the fractional derivatives (left and right Riemann- Liouville fractional derivatives) are discussed. These definitions are used in the Hamiltonian formulation and the solution of examples leading to the equations of motion of the fractional order.

The left Riemann-Liouville fractional derivative (LRLFD) reads as[25]:

$$ {}_aD_x^{\alpha} f(x) = \frac{1}{\Gamma(n-\alpha)} \left(\frac{d}{dx}\right)^n \int_a^x \frac{f(\tau)}{(x-\tau)^{\alpha-n+1}} d\tau . \quad (1)$$

The right Riemann-Liouville fractional derivative (RRLFD) reads as[25]:

$$ {}_xD_b^{\alpha} f(x) = \frac{1}{\Gamma(n-\alpha)} \left(-\frac{d}{dx}\right)^n \int_x^b \frac{f(\tau)}{(\tau-x)^{\alpha-n+1}} d\tau . \quad (2)$$

Here $\alpha$ is the order of the derivative such that $n-1 \le \alpha \le n$ and is not equal to zero. If $\alpha$ is an integer, these derivatives are defined in the usual sense, i.e.,

$$ {}_aD_x^{\alpha} f(x) = \left(\frac{d}{dx}\right)^{\alpha} f(x) ;$$

$$ {}_xD_b^{\alpha} f(x) = \left(-\frac{d}{dx}\right)^{\alpha} f(x) ; \ \alpha = 1,2,.... \quad (3)$$

## 3. Fractional form of Euler-Lagrangian Equation in terms of Functional derivative of the full Lagrangian L

We start our formalism by taking the Lagrangian density to be a function of field amplitude $\psi$ and its fractional derivatives with respect to space and time as:

$$ \pounds = \pounds(\psi, {}_aD_x^{\alpha}\psi, {}_xD_b^{\beta}\psi, {}_aD_t^{\alpha}\psi, {}_tD_b^{\beta}\psi, t) .$$
(4)

Euler-Lagrange equation for such Lagrangian density in fractional form can be given as[26]:

$$ \frac{\partial \pounds}{\partial \psi} + {}_xD_b^{\alpha} \frac{\partial \pounds}{\partial\, {}_aD_x^{\alpha}\psi} + {}_aD_x^{\beta} \frac{\partial \pounds}{\partial\, {}_xD_b^{\beta}\psi} + {}_tD_b^{\alpha} \frac{\partial \pounds}{\partial\, {}_aD_t^{\alpha}\psi} + {}_aD_t^{\beta} \frac{\partial \pounds}{\partial\, {}_tD_b^{\beta}\psi} = 0 .$$
(5)

Now we can write the full Lagrangian $L$ as:

$$L = \int \mathcal{L}\, d^3 r . \tag{6}$$

Using the variational principle, we can write:

$$\delta \int L dt = \delta \int \int \mathcal{L} d^3 r dt = \int \int (\delta \mathcal{L}) dt d^3 r = 0 . \tag{7}$$

Using Eq. (4), the variation of is:

$$\delta \mathcal{L} = \frac{\partial \mathcal{L}}{\partial \psi} \delta \psi + \frac{\partial \mathcal{L}}{\partial\, {}_a D_x^\alpha \psi} \delta ({}_a D_x^\alpha \psi) + \frac{\partial \mathcal{L}}{\partial\, {}_x D_b^\beta \psi} \delta ({}_x D_b^\beta \psi) + \frac{\partial \ell}{\partial\, {}_a D_t^\alpha \psi} \delta ({}_a D_t^\alpha \psi) + \frac{\partial \mathcal{L}}{\partial\, {}_t D_b^\beta \psi} \delta ({}_t D_b^\beta \psi) = 0 . \tag{8}$$

Substituting Eq. (8) into Eq. (7), and using the following commutation relation,

$$\frac{\partial \mathcal{L}}{\partial\, {}_a D_x^\alpha \psi} \delta ({}_a D_x^\alpha \psi) = \frac{\partial \mathcal{L}}{\partial\, {}_a D_x^\alpha \psi}\, {}_a D_x^\alpha (\delta \psi) . \tag{9}$$

we get,

$$\int \int [ \frac{\partial \mathcal{L}}{\partial \psi} (\delta \psi) + \frac{\partial \mathcal{L}}{\partial\, {}_a D_x^\alpha \psi}\, {}_a D_x^\alpha (\delta \psi) + \frac{\partial \ell}{\partial\, {}_x D_b^\beta \psi}\, {}_x D_b^\beta (\delta \psi) + \frac{\partial \mathcal{L}}{\partial\, {}_a D_t^\alpha \psi} \delta ({}_a D_t^\alpha \psi) + \frac{\partial \mathcal{L}}{\partial\, {}_t D_b^\beta \psi} \delta ({}_t D_b^\beta \psi) ]\, d^3 r\, dt = 0 . \tag{10}$$

Integrating by parts the second and the third terms in Eq. (10) with respect to space, we get:

$$0 = \int dt \int [ \frac{\partial \mathcal{L}}{\partial \psi} + {}_x D_b^\alpha \frac{\partial \mathcal{L}}{\partial\, {}_a D_x^\alpha \psi} + {}_a D_x^\beta \frac{\partial \ell}{\partial\, {}_x D_b^\beta \psi} ] d\tau \delta \psi + \int dt \int [ \frac{\partial \mathcal{L}}{\partial\, {}_a D_t^\alpha \psi} \delta ({}_a D_t^\alpha \psi) + \frac{\partial \mathcal{L}}{\partial\, {}_t D_b^\beta \psi} \delta ({}_t D_b^\beta \psi) ] d\tau . \tag{11}$$

Now, we can take the integration over space $d\tau$ in the previous equation and convert it into summation, thus:

$$\sum_i \left[ \frac{\partial \mathcal{L}}{\partial \psi} + {}_x D_b^\alpha \frac{\partial \mathcal{L}}{\partial\, {}_a D_x^\alpha \psi} + {}_a D_x^\beta \frac{\partial \mathcal{L}}{\partial\, {}_x D_b^\beta \psi} \right]_i \delta \psi_i \delta \tau_i + \sum_i \left[ \frac{\partial \mathcal{L}}{\partial\, {}_a D_t^\alpha \psi} \right]_i \delta ({}_a D_t^\alpha \psi_i) \delta \tau_i + + \sum_i \left[ \frac{\partial \mathcal{L}}{\partial\, {}_t D_b^\beta \psi} \right]_i \delta ({}_t D_b^\beta \psi_i) \delta \tau_i = 0 . \tag{12}$$

We can write Eq. (12) in terms of Lagrangian density as:

$$\sum_i [\delta \mathcal{L}]_i \delta\tau_i = 0 . \qquad (13)$$

Where the left hand side in Eqs.(12 and 13) represents the variation of $L$ (i.e. $\delta L$) which is now produced by independent variations in $\delta\psi_i$, $\delta({}_aD_t^\alpha \psi_i)$ and $\delta({}_tD_b^\beta \psi_i)$. Suppose now that all $\delta\psi_i$ , $\delta({}_aD_t^\alpha \psi_i)$ and $\delta({}_tD_b^\beta \psi_i)$ are zeros except for a particular $\delta\psi_j$. It is natural to define the functional derivative of the full Lagrangian ($\partial L$) with respect to $\psi$, $({}_aD_t^\alpha \psi)$ and $({}_tD_b^\beta \psi)$ for a point in the j-th cell to the ratio of $\delta L$ to $\delta\psi_j$ [27].

$$\frac{\partial L}{\partial \psi} = \lim_{\delta\tau_j \to 0} \frac{\delta L}{\delta\psi_j \, \delta\tau_j} . \qquad (14)$$

Using Eq. (12), and note that the left hand side represents **$\delta L$,** we get:

$$\frac{\partial L}{\partial \psi} = \frac{\partial \mathcal{L}}{\partial \psi} + {}_xD_b^\alpha \left[ \frac{\partial \mathcal{L}}{\partial\, {}_aD_x^\alpha \psi} \right] + {}_aD_x^\beta \left[ \frac{\partial \mathcal{L}}{\partial\, {}_xD_b^\beta \psi} \right] . \qquad (15a)$$

$$\frac{\partial L}{\partial\, {}_aD_t^\alpha \psi} = \lim_{\delta\tau_j \to 0} \frac{\delta L}{\delta\tau_j \, \delta({}_aD_t^\alpha \psi_j)} = \frac{\partial \mathcal{L}}{\partial\, {}_aD_t^\alpha \psi} . \qquad (15b)$$

$$\frac{\partial L}{\partial\, {}_tD_b^\beta \psi} = \lim_{\delta\tau_j \to 0} \frac{\delta L}{\delta\tau_j \delta({}_tD_b^\beta \psi_j)} = \frac{\partial \mathcal{L}}{\partial\, {}_tD_b^\beta \psi} . \qquad (15c)$$

Now, using Eq. (15) we can rewrite Eq. (5) Euler- Lagrange equation in terms of the full Lagrangian $L$ using functional derivative in fractional form:

$$\frac{\partial L}{\partial \psi} + {}_tD_b^\alpha \left[ \frac{\partial L}{\partial_a D_t^\alpha \psi} \right] + {}_aD_t^\beta \left[ \frac{\partial L}{\partial_t D_b^\beta \psi} \right] = 0 . \qquad (16)$$

It is worth mentioning that for $\alpha, \beta \to 1$, Eq. (16) reduces to the usual Euler- Lagrange equation for the classical fields[27]. With the help of Eqs. (12 and 16), we can write the variation of full Lagrangian in terms of functional derivatives and variations of $\psi$, ${}_aD_t^\alpha \psi$ and ${}_tD_b^\beta \psi$ as:

$$\delta L = \int \left[ \frac{\partial L}{\partial \psi} (\delta\psi) + \frac{\partial L}{\partial_a D_t^\alpha \psi} \delta({}_aD_t^\alpha \psi) + \frac{\partial L}{\partial_t D_b^\beta \psi} \delta({}_tD_b^\beta \psi) \right] d^3 r . \qquad (17)$$

## 4. Fractional form of Euler-Lagrange Equation in terms of Momentum Density

The right side fractional form of momentum can be written as [27]:

$$P_j^a = \frac{\delta L}{\delta\,{}_aD_t^\alpha \psi_j}. \tag{18a}$$

Using Esq. (12 and 15b) we get:

$$P_j^a = \frac{\partial \mathcal{L}}{\partial\,{}_aD_t^\alpha \psi_j}\delta\tau_j = \frac{\partial L}{\partial\,{}_aD_t^\alpha \psi_j}\delta\tau_j. \tag{18b}$$

From Eq. (18b), we can define the right side form of momentum density $\pi_\alpha$ as:

$$(\pi_\alpha)_j = \frac{\partial L}{\partial\,{}_aD_t^\alpha \psi_j} = \frac{\partial \mathcal{L}}{\partial\,{}_aD_t^\alpha \psi_j}. \tag{19}$$

Now, taking the left fractional derivative for Eq. (19), one gets:

$${}_tD_b^\alpha (\pi_\alpha)_j = {}_tD_b^\alpha\left[\frac{\partial L}{\partial\,{}_aD_t^\alpha \psi_j}\right]. \tag{20}$$

Repeating same steps above for left side fractional form of momentum density $\pi_\beta$, we get:

$$(\pi_\beta)_j = \frac{\partial L}{\partial\,{}_tD_b^\beta \psi_j} = \frac{\partial \mathcal{L}}{\partial\,{}_tD_b^\beta \psi_j}. \tag{21}$$

Now, taking the right fractional derivative for Eq. (21), one gets:

$${}_aD_t^\beta (\pi_\beta)_j = {}_aD_t^\beta\left[\frac{\partial L}{\partial\,{}_tD_b^\beta \psi_j}\right]. \tag{22}$$

Now, substituting Eqs. (20 and 22) into Eq. (16), we get:

$$\frac{\partial L}{\partial \psi} = -[{}_aD_t^\beta \pi_\beta + {}_tD_b^\alpha \pi_\alpha]. \tag{23}$$

The above equation represents the fractional form of Euler- Lagrange equation in terms of momentum density and the functional derivative of full Lagrangian.

## 5. Equations of Motion in terms of Hamiltonian Density in Fractional Form

We start by the general definition of the Hamiltonian density $h$ in fractional form as:

$$\mathcal{H} = \pi_\alpha \, {}_aD_t^\alpha \psi + \pi_\alpha^* \, {}_aD_t^\alpha \psi^* + \pi_\beta \, {}_tD_b^\beta \psi + \pi_\beta^* \, {}_tD_b^\beta \psi^* - \mathcal{L}. \tag{24}$$

Full Hamiltonian $H$ can be also written in terms of Hamiltonian density $\mathcal{H}$ as:

$$H = \sum_i \mathcal{H}_i \delta\tau_i. \tag{25}$$

Substituting Eqs. (24) into Eq. (25), one gets:

$$H = \sum_i \left[ (\pi_\alpha)_i ({}_aD_t^\alpha \psi_i) + (\pi_\alpha^*)_i ({}_aD_t^\alpha \psi_i^*) + (\pi_\beta)_i ({}_tD_b^\beta \psi_i) + (\pi_\beta^*)_i ({}_tD_b^\beta \psi_i^*) \right] \delta\tau_i - \sum_i \mathcal{L}_i \delta\tau_i. \tag{26}$$

In continuous form, we can write Eq. (26) as follows:

$$H = \int \left[ (\pi_\alpha)({}_aD_t^\alpha \psi) + (\pi_\alpha^*)({}_aD_t^\alpha \psi^*) + (\pi_\beta)({}_tD_b^\beta \psi) + (\pi_\beta^*)({}_tD_b^\beta \psi^*) \right] d^3r - \int \mathcal{L} d^3r. \tag{27}$$

Taking the variation of H, using Eqs. (17 and 23), see appendix A, we get:

$$\delta H = \int [({}_aD_t^\beta \pi_\beta + {}_tD_b^\alpha \pi_\alpha)\delta\psi + ({}_aD_t^\beta \pi_\beta^* + {}_tD_b^\alpha \pi_\alpha^*)\delta\psi^* + ({}_aD_t^\alpha \psi)\delta\pi_\alpha + ({}_aD_t^\alpha \psi^*)\delta\pi_\alpha^* + ({}_tD_b^\beta \psi)\delta\pi_\beta + ({}_tD_b^\beta \psi^*)\delta\pi_\beta^*] d^3r. \tag{28}$$

By analogy with the variation in $L$ (i.e: Eq. (17)), we can write the variation of full Hamiltonian produced by variations of independent variables in terms of functional derivative as follows in cases 1 and 2.

***Case 1:*** All variables are independent ($\psi, \psi^*, \pi_\alpha, \pi_\beta, \pi_\alpha{}^*, and \quad \pi_\beta{}^*$)

$$\delta H = \int [\frac{\partial H}{\partial \psi}\delta\psi + \frac{\partial H}{\partial \psi^*}\delta\psi^* + \frac{\partial H}{\partial \pi_\alpha}\delta\pi_\alpha + \frac{\partial H}{\partial \pi_\alpha^*}\delta\pi_\alpha^* + \frac{\partial H}{\partial \pi_\beta}\delta\pi_\beta + \frac{\partial H}{\partial \pi_\beta^*}\delta\pi_\beta^*]d^3r. \tag{29}$$

Comparing Eq. (29) with Eq. (28), we get the separate equations of motion in terms of full Hamiltonian as:

$$\frac{\partial H}{\partial \psi} = {}_aD_t^\beta \pi_\beta + {}_tD_b^\alpha \pi_\alpha; \qquad \frac{\partial H}{\partial \psi^*} = {}_aD_t^\beta \pi_\beta^* + {}_tD_b^\alpha \pi_\alpha^*. \tag{30a}$$

$$\frac{\partial H}{\partial \pi_\alpha} = {}_aD_t^\alpha \psi \;\; ; \;\; \frac{\partial H}{\partial \pi_\alpha^*} = {}_aD_t^\alpha \psi^* \;\; ; \;\; \frac{\partial H}{\partial \pi_\beta} = {}_tD_b^\beta \psi \;\; ; \;\; \frac{\partial H}{\partial \pi_\beta^*} = {}_tD_b^\beta \psi^*. \tag{30b}$$

By analogy with Eq. (15a) for functional derivative of full Lagranian in terms of fractional derivative of Lagrangian density, we can simply define the functional derivative of $H$ in terms of fractional derivative of Hamiltonian density with respect to the general variable field $\phi$ as [27]:

$$\frac{\partial H}{\partial \varphi} = \frac{\partial \mathcal{H}}{\partial \varphi} + {}_xD_b^\alpha \frac{\partial \mathcal{H}}{\partial\, {}_aD_x^\alpha \varphi} + {}_aD_x^\beta \frac{\partial \mathcal{H}}{\partial\, {}_xD_b^\beta \varphi} \tag{31}$$

Using the definition given in Eq. (31) above, we can rewrite equations of motion (i.e. Eqs. (30)) in terms of Hamiltonian density such that:

$$\frac{\partial \mathcal{H}}{\partial \psi} + {}_xD_b^\alpha \frac{\partial \mathcal{H}}{\partial\, {}_aD_x^\alpha \psi} + {}_aD_x^\beta \frac{\partial \mathcal{H}}{\partial\, {}_xD_b^\beta \psi} = {}_aD_t^\beta \pi_\beta + {}_tD_b^\alpha \pi_\alpha. \tag{32a}$$

$$\frac{\partial \mathcal{H}}{\partial \psi^*} + {}_xD_b^\alpha \frac{\partial \mathcal{H}}{\partial\, {}_aD_x^\alpha \psi^*} + {}_aD_x^\beta \frac{\partial \mathcal{H}}{\partial\, {}_xD_b^\beta \psi^*} = {}_aD_t^\beta \pi_\beta^* + {}_tD_b^\alpha \pi_\alpha^* \;. \tag{32b}$$

$$\frac{\partial \mathcal{H}}{\partial \pi_\alpha} + {}_xD_b^\alpha \frac{\partial \mathcal{H}}{\partial\, {}_aD_x^\alpha \pi_\alpha} + {}_aD_x^\beta \frac{\partial \mathcal{H}}{\partial\, {}_xD_b^\beta \pi_\alpha} = {}_aD_t^\alpha \psi. \tag{32c}$$

$$\frac{\partial \mathcal{H}}{\partial \pi_\alpha^*} + {}_xD_b^\alpha \frac{\partial \mathcal{H}}{\partial\, {}_aD_x^\alpha \pi_\alpha^*} + {}_aD_x^\beta \frac{\partial \mathcal{H}}{\partial\, {}_xD_b^\beta \pi_\alpha^*} = {}_aD_t^\alpha \psi^*. \tag{32d}$$

$$\frac{\partial \mathcal{H}}{\partial \pi_\beta} + {}_xD_b^\alpha \frac{\partial \mathcal{H}}{\partial\, {}_aD_x^\alpha \pi_\beta} + {}_aD_x^\beta \frac{\partial \mathcal{H}}{\partial\, {}_xD_b^\beta \pi_\beta} = {}_tD_b^\beta \psi. \tag{32e}$$

$$\frac{\partial \mathcal{H}}{\partial \pi_\beta^*} + {}_xD_b^\alpha \frac{\partial \mathcal{H}}{\partial\, {}_aD_x^\alpha \pi_\beta^*} + {}_aD_x^\beta \frac{\partial \mathcal{H}}{\partial\, {}_xD_b^\beta \pi_\beta^*} = {}_tD_b^\beta \psi^*. \tag{32f}$$

In many cases, we take $\pi_\beta = 0$ because we define (in the Lagrangian density and the Hamiltonian density) the time derivative in the right side as ${}_aD_t^\alpha \psi$, so that $\pi_\beta = \frac{\partial \mathcal{L}}{\partial_t D_b^\beta \psi} = 0$. Therefore take $\pi_\beta = 0$, and $\pi_\beta^* = 0$.

***Case 2:*** $\pi_\alpha$ depends on $(\psi,\ or\ \psi^*)$ and $\pi_\alpha^*$ depends on $(\psi,\ or\ \psi^*)$; so that we take the variation just only for independent variables $\psi$, and $\psi^*$. Thus Eq. (31) can be written as:

$$\delta H = \int \left[ \frac{\partial H}{\partial \psi} \delta\psi + \frac{\partial H}{\partial \psi^*} \delta\psi^* \right] d^3 r . \qquad (33)$$

To state the equations of motion from Eq. (29), let us define $\pi_\alpha$ and $\pi_\alpha^*$ in a general case $\pi_\alpha = g(\psi, \psi^*)$ and $\pi_\alpha^* = f(\psi, \psi^*)$. So that, we can write their variations as:

$$\delta\pi_\alpha = \frac{\partial g}{\partial \psi} \delta\psi + \frac{\partial g}{\partial \psi^*} \delta\psi^* . \qquad (34)$$

$$\delta\pi_\alpha^* = \frac{\partial f}{\partial \psi} \delta\psi + \frac{\partial f}{\partial \psi^*} \delta\psi^* . \qquad (35)$$

Now, substituting Eqs. (34 and 35) into Eq.(29) , and comparing with Eq. (33) , we get the general equations of the Hamiltonian density for this case:

$$\frac{\partial \mathcal{H}}{\partial \psi} + {}_xD_b^\alpha \frac{\partial \mathcal{H}}{\partial\, {}_aD_x^\alpha \psi} + {}_aD_x^\beta \frac{\partial h}{\partial\, {}_xD_b^\beta \psi} = {}_tD_b^\alpha \pi_\alpha + \frac{\partial g}{\partial \psi}\, {}_aD_t^\alpha \psi + \frac{\partial f}{\partial \psi}\, {}_aD_t^\alpha \psi^* . \qquad (36)$$

$$\frac{\partial \mathcal{H}}{\partial \psi^*} + {}_xD_b^\alpha \frac{\partial \mathcal{H}}{\partial\, {}_aD_x^\alpha \psi^*} + {}_aD_x^\beta \frac{\partial h}{\partial\, {}_xD_b^\beta \psi^*} = {}_tD_b^\alpha \pi_\alpha^* + \frac{\partial g}{\partial \psi^*}\, {}_aD_t^\alpha \psi + \frac{\partial f}{\partial \psi^*}\, {}_aD_t^\alpha \psi^* . \qquad (37)$$

## 6. Examples

In this section, we study two examples as applications on the formalism presented above.

***Example 1:*** Schrödinger Equation

*Lagrangian density in fractional form:*

$$\mathcal{L} = -\frac{\hbar^2}{2m}({}_aD_x^\alpha\psi^*)({}_aD_x^\alpha\psi) - \frac{\hbar}{2i}[\psi^*\left({}_aD_t^\alpha\psi\right) - \left({}_aD_t^\alpha\psi^*\right)\psi] - \psi^* V\psi \,. \tag{38}$$

Applying Euler-Lagrange equation (Eq. (5)) with respect to $\psi^*$, we get:

$$\frac{\hbar^2}{2m}{}_xD_b^\alpha({}_aD_x^\alpha\psi) + V\psi = \frac{\hbar i}{2}({}_aD_t^\alpha\psi - {}_tD_b^\alpha\psi) \,. \tag{39}$$

Now we want to derive Eq.(39) using the Hamiltonian density equations of motion. First we determine $\pi_\alpha, \pi_\alpha^*, \pi_\beta$ and $\pi_\beta^*$ using Eqs.(19 and 21):

$$\pi_\alpha = \frac{\partial\mathcal{L}}{\partial\, {}_aD_t^\alpha\psi} = \frac{-\hbar}{2i}\psi^* ; \qquad \pi_\alpha^* = \frac{\partial\mathcal{L}}{\partial\, {}_aD_t^\alpha\psi^*} = \frac{\hbar}{2i}\psi \,. \tag{40}$$

$$\pi_\beta = \frac{\partial\mathcal{L}}{\partial\, {}_tD_b^\beta\psi} = 0 ; \qquad \pi_\beta^* = \frac{\partial\mathcal{L}}{\partial\, {}_tD_b^\beta\psi^*} = 0 \,. \tag{41}$$

Then, using Eq.(24), the Hamiltonian density can be written as:

$$\mathcal{H} = \frac{\hbar^2}{2m}({}_aD_x^\alpha\psi^*)({}_aD_x^\alpha\psi) + \psi^* V\psi \,. \tag{42}$$

Now, because $\pi_\alpha$ and $\pi_\alpha^*$ are variables dependent of $\psi^*$ , and $\psi$, respectively, we have to use equations of motion for case2. Applying Eq. (37), we get:

$$V\psi + \frac{\hbar^2}{2m}{}_xD_b^\alpha({}_aD_x^\alpha\psi) = \frac{\hbar}{2i}({}_aD_t^\alpha\psi - {}_tD_b^\alpha\psi) \,. \tag{43}$$

The above equation is exactly the same as the equation that has been derived by Euler- Lagrange (Eq. (39)) in fractional form.

If $\alpha = \beta = 1$, then Eq. (39) and Eq. (43) become:

$$\frac{-\hbar^2}{2m}\nabla^2\psi + V\psi = i\hbar\frac{\partial\psi}{\partial t} \,. \tag{44}$$

This is the known Schrödinger equation.
If we do not consider the dependency of $\pi_\alpha$ and $\pi_\alpha^*$ on $\psi^*$ , and $\psi$, respectively, and apply Eq. (32b) in case1, then we get:

$$\frac{\hbar^2}{2m}\,{}_xD_b^\alpha({}_aD_x^\alpha\psi)+V\psi=\frac{-\hbar}{2i}({}_tD_b^\alpha\psi)\,. \tag{45}$$

If $\alpha=\beta=1$, then Eq. (45) becomes:

$$\frac{-\hbar^2}{2m}\nabla^2\psi+V\psi=\frac{i\hbar}{2}\frac{\partial\psi}{\partial t}\,. \tag{46}$$

This is not equivalent to Schrödinger equation given by Eq.(44).This means that equations of motion mentioned in case1 do not represent the general case for equations of motion in terms of Hamiltonian density as in ref. 24. But they only represent a special case for independency of $\psi,\psi^*,\pi_\alpha,\pi_\beta,\pi_\alpha^*, and\ \ \pi_\beta^*$ .

***Example 2:*** Dirac Equation

The Lagrangian density is[28]*:*

$$\mathcal{L}=\frac{\hbar c}{2i}\left[(\vec{\nabla}\psi^*).\vec{\alpha}\psi-\psi^*\vec{\alpha}.(\vec{\nabla}\psi)\right]+\frac{\hbar}{2i}\left[\frac{\partial\psi^*}{\partial t}\psi-\psi^*\frac{\partial\psi}{\partial t})\right]-e\psi^*(\vec{\alpha}.\vec{A})\psi+$$
$$ec\psi^*\phi\psi-mc^2\psi^*\alpha_o\psi\,. \tag{47}$$

Where $\phi$ and $\vec{A}$ are electromagnetic potential.

$e$ and $m$ are electron's charge and mass respectively.

$\psi$ and $\psi^*$ have four components $(i.e:\ \psi=(\psi_1,\psi_2,\psi_3,\psi_4), and\ \ \psi^*=(\psi_1^*,\psi_2^*,\psi_3^*,\psi_4^*)$ with four spin vectors.

$\alpha$ an operator with four components (i.e: $\vec{\alpha}\,,\alpha_o$)

With $\vec{\alpha}=\alpha_x\hat{i}+\alpha_y\hat{j}+\alpha_z\hat{k}$

We apply our equations for $\psi$ and $\psi^*$ in general and then we can generalize them for any given $\psi_j$ and $\psi_j^*$.

Lagrangian density in fractional form is:

$$\mathcal{L}=\frac{\hbar c}{2i}\left[({}_aD_{x_j}^\alpha\psi^*)\alpha_j\psi-\psi^*\alpha_j({}_aD_{x_j}^\alpha\psi)\right]-\frac{\hbar}{2i}\left[({}_aD_t^\alpha\psi^*)\psi-\psi^*({}_aD_t^\alpha\psi)\right]-e\psi^*(\vec{\alpha}.\vec{A})\psi+$$
$$ec\psi^*\phi\psi-mc^2\psi^*\alpha_o\psi\,. \tag{48}$$

Here the subscript (j) means sum over (x,y,z).

Equations of motion, using Euler- Lagrange equation Eq.(5), by taking the derivative with respect to $\psi^*$, we get:

$$mc\alpha_o\psi+\frac{\hbar}{2i}\alpha_j\left[{}_aD^{\alpha}_{x_j}\psi-{}_{x_j}D^{\alpha}_b\psi+\frac{e}{c}A_j\,\psi\right]+\frac{\hbar}{2ic}\left[{}_aD^{\alpha}_t\psi-\psi^*{}_tD^{\alpha}_b\psi\right]-e\phi\psi=0\,. \qquad (49)$$

For $\alpha=1$ we get the real equation of motion:

$$mc\alpha_o\psi+\vec{\alpha}.\left(\frac{\hbar}{i}\vec{\nabla}\,\psi+\frac{e}{c}\vec{A}\,\psi\right)+\left[\frac{\hbar}{ic}\frac{\partial\psi}{\partial t}-e\phi\psi\right]=0\;. \qquad (50)$$

We can determine $\pi_\alpha\ and\ \pi^*_\alpha$ such that:

$$\pi_\alpha=\frac{\partial\mathcal{L}}{\partial\,{}_aD^{\alpha}_t\psi}=\frac{-\hbar}{2i}\psi^*\;. \qquad (51)$$

$$\pi^*_\alpha=\frac{\partial\mathcal{L}}{\partial\,{}_aD^{\alpha}_t\psi^*}=\frac{\hbar}{2i}\psi\,. \qquad (52)$$

Substituting Eqs.(51 and 52) into Eq. (21), the Hamiltonian density can be written as:

$$\mathcal{H}=\frac{-\hbar c}{2i}\left[({}_aD^{\alpha}_{x_j}\psi^*)\alpha_j\,\psi-\psi^*\alpha_j({}_aD^{\alpha}_{x_j}\psi^*)\right]+e\psi^*(\vec{\alpha}.\vec{A})\psi-ec\psi^*\varphi\psi+mc^2\psi^*\alpha_o\psi\;. \quad (53)$$

We note from Eqs(51 and 52) that $\pi_\alpha\ and\ \pi^*_\alpha$ are depending on $\psi^*\ and\ \psi$, respectively, so that we have to use equations of motion for case2. Applying Eq. (37) with respect to $\psi^*$, we get:

$$mc\alpha_o\psi+\frac{\hbar c}{2i}\alpha_j\left[{}_aD^{\alpha}_{x_j}\psi^*-{}_{x_j}D^{\alpha}_b\psi+\frac{e}{c}A_j\,\psi\right]+\frac{\hbar}{2ic}\left[{}_aD^{\alpha}_t\psi-{}_tD^{\alpha}_b\psi\right]-e\phi\psi=0. \qquad (54)$$

This result is the same as that obtained by Euler- Lagrange, see Eq. (49). For $\alpha=1$, we get Eq. (50) in real space.

If we do not consider the dependency condition of $\pi_\alpha$ and $\pi^*_\alpha$ on $\psi^*\ and\ \psi$, respectively, and apply Eq. (32b) for case1, we get:

$$mc\alpha_o\psi+\frac{\hbar c}{2i}\alpha_j\left[{}_aD^{\alpha}_{x_j}\psi-{}_{x_j}D^{\alpha}_b\psi+\frac{e}{c}A_j\psi\right]+\frac{-\hbar}{2ic}\,{}_tD^{\alpha}_b\psi-e\phi\psi=0\,. \qquad (55)$$

Eq.(55) is not equivalent to Eq.(50), that is derived from Euler- Lagrange equation. By considering $\alpha = 1$ , we get:

$$mc\psi\alpha_o + \vec{\alpha}.(\frac{\hbar}{i}\vec{\nabla}\psi + \frac{e}{c}\vec{A}\psi) + \frac{\hbar}{2ic}\frac{\partial\psi}{\partial t} - e\phi\psi = 0\,. \qquad (56)$$

This equation differs from Eq.(50) by a factor of $\frac{1}{2}$ appearing in the time derivative term of $\frac{\hbar}{2ic}\frac{\partial\psi}{\partial t}$.

## 7. Conclusion

We constructed the Hamiltonian formulation of continuous field systems. Our results are the same as those derived by using the formulation of Euler- Lagrange. Two cases are considered: the conjugate momenta are field dependent or field independent. As special cases, for derivatives of integer orders only, the results of the equations of motion are found in agreement with the Lagrangian formulation of continuous systems.

**Appendix A**
**Variation of full Hamiltonian**

We can rewrite Eq. (27) as:

$$H = \int \left[ \pi_\alpha \, {}_aD_t^\alpha \psi + \pi_\alpha^* \, {}_aD_t^\alpha \psi^* + \pi_\beta \, {}_tD_b^\beta \psi + \pi_\beta^* \, {}_tD_b^\beta \psi^* \right] d^3r - L. \quad \text{(A1)}$$

Now, take the variation of H, , we get:

$$\delta H = \int \delta(\pi_\alpha \, {}_aD_t^\alpha \psi + \pi_\beta \, {}_tD_b^\beta \psi) d^3r + \int \delta(\pi_\alpha^* \, {}_aD_t^\alpha \psi^* + \pi_\beta^* \, {}_tD_b^\beta \psi^*) d^3r - \delta L. \quad \text{(A2)}$$

Using Eq. (19), Eq. (21) and Eq. (23), we rewrite the variation of full Lagrangian given by Eq. (17) as:

$$\delta L = \int \{ -({}_aD_t^\beta \pi_\beta + {}_tD_b^\alpha \pi_\alpha) \delta\psi + \pi_\alpha \, \delta({}_aD_t^\alpha \psi) + \pi_\beta \, \delta({}_tD_b^\beta \psi) \} d^3r. \quad \text{(A3)}$$

The above equation can be arranged as:

$$\delta L = \int \{ -({}_aD_t^\beta \pi_\beta + {}_tD_b^\alpha \pi_\alpha) \delta\psi + \delta(\pi_\alpha \, {}_aD_t^\alpha \psi + \pi_\beta \, {}_tD_b^\beta \psi) - {}_aD_t^\alpha \psi \, \delta\pi_\alpha - {}_tD_b^\beta \psi \, \delta\pi_\beta \} d^3r.$$

(A4)

Substituting Eq. (A4) into Eq. (A2), one gets:

$$\delta H = \int \{ ({}_aD_t^\beta \pi_\beta + {}_tD_b^\alpha \pi_\alpha) \delta\psi + {}_aD_t^\alpha \psi \, \delta\pi_\alpha + {}_aD_t^\alpha \psi^* \, \delta\pi_\alpha^* + {}_tD_b^\beta \psi \, \delta\pi_\beta + {}_tD_b^\beta \psi^* \, \delta\pi_\beta^* + \pi_\alpha^* \, {}_aD_t^\alpha (\delta\psi^*) + \pi_\beta^* \, {}_tD_b^\beta (\delta\psi^*) \} d^3r.$$

(A5)

Taking the time integration of $\delta H$ , we get:

$$\int \delta H dt = \int dt \int d^3r \{ ({}_aD_t^\beta \pi_\beta + {}_tD_b^\alpha \pi_\alpha) \delta\psi + {}_aD_t^\alpha \psi \, \delta\pi_\alpha + {}_aD_t^\alpha \psi^* \, \delta\pi_\alpha^* + {}_tD_b^\beta \psi \, \delta\pi_\beta + {}_tD_b^\beta \psi^* \, \delta\pi_\beta^* + \pi_\alpha^* \, {}_aD_t^\alpha (\delta\psi^*) + \pi_\beta^* \, {}_tD_b^\beta (\delta\psi^*) \}.$$

(A6)

Integrate (by parts) the last two terms in the above equation then separate integrals over time and integrals over space to get:

$$\delta H = \int [ ({}_aD_t^\beta \pi_\beta + {}_tD_b^\alpha \pi_\alpha) \delta\psi + ({}_aD_t^\beta \pi_\beta^* + {}_tD_b^\alpha \pi_\alpha^*) \delta\psi^* + ({}_aD_t^\alpha \psi) \delta\pi_\alpha + ({}_aD_t^\alpha \psi^*) \delta\pi_\alpha^* + ({}_tD_b^\beta \psi) \delta\pi_\beta + ({}_tD_b^\beta \psi^*) \delta\pi_\beta^* ] d^3r.$$

(A7)